\documentclass[aps,preprint,preprintnumbers,amsmath,amssymb,nofootinbib]{revtex4}
\usepackage{amssymb}
\usepackage{amsmath}
\usepackage{bm}

\begin{document}
\title{Fierz-Pauli equation for massive gravitons from Induced Matter theory of gravity}
\author{$^{1,2}$ Mauricio Bellini \footnote{ E-mail address:
mbellini@mdp.edu.ar, mbellini@conicet.gov.ar} }
\address{$^1$ Departamento de F\'isica, Facultad de Ciencias Exactas y
Naturales, Universidad Nacional de Mar del Plata, Funes 3350,
C.P. 7600, Mar del Plata, Argentina.\\  \\
$^2$ Instituto de Investigaciones F\'{\i}sicas de Mar del Plata (IFIMAR), \\
Consejo acional de Investigaciones Cient\'ificas y T\'ecnicas
(CONICET), Argentina.}

\begin{abstract}
Starting with a 5D physical vacuum described by a 5D Ricci-flat
background metric, we study the emergence of gravitational waves
(GW) from the Induce Matter (IM) theory of gravity.  We obtain the
equation of motion for GW on an 4D curved spacetime which has the
form of a Fierz-Pauli one. In our model the mass of gravitons
$m_g$ is induced by a static foliation on the noncompact
space-like extra dimension and the source-term is originated in
the interaction of the GW with the induced connections of the
background 5D metric. Here, relies the main difference of this
formalism with the original Fierz-Pauli one.
\end{abstract}

\maketitle

\section{Introduction}

Although standard general relativity (GR)\cite{ei} achieved great
success and withstood many experimental tests, it also displayed
many shortcomings and flaws which today make theoreticians
question whether it is the definitive theory of gravity
[see\cite{gr} and references therein]. It is well known that GR is
very difficult to quantize. This fact rules out the possibility of
treating gravitation like other quantum theories, and precludes
the unification of gravity with other interactions. At the present
time, it is not possible to realize a consistent quantum gravity
theory which leads to the unification of gravitation with the
other forces. In particular, the theory of gravitational waves
(GW) is a rich subject that brings together different domains such
as general relativity, field theory, astrophysics and cosmology.
At present various gravitational-wave detectors, after decades of
developments, have reached a sensitivity where there are
significant chances of detection, and future improvements are
expected to lead, in a few years, to advanced detectors with even
better sensitivities\cite{maggiore}. The tensor perturbations of
the metric $h_{\mu\nu}$ propagate and can affect the background
space-time $\bar{g}_{\mu\nu}$\cite{bard}. Since the discovery of
the CMB electromagnetic radiation we know that its spectrum is a
perfect black-body. This background is, to a first order
approximation, isotropic. There are good reasons to expect that
the Universe is permeated also by a stochastic background of GWs
generated in the early universe. Furthermore, a stochastic
background can also emerge from the incoherent superposition of a
large number of astrophysical sources, too weak to be detected
separately, and such that the number of sources that contribute to
each frequency bin is much larger than one. The inflationary
theory is one of the better candidates to describe the early
stages of the accelerated expansion of the universe\cite{guth}.
This theory can be recovered from from the 5D IM
one\cite{bell,gwi}.

In this letter we study the emergence of gravitational waves (GW)
from the Induce Matter (IM) theory of gravity\cite{im}. We shall
investigate how we can to obtain the equation that describes the
evolution of GW from a 5D vacuum state, which is defined from a
Ricci-flat spacetime. We shall restrict to canonical
metrics\cite{wo} which are, at least, 5D Ricci-flat, on which we
shall define a 5D vacuum state.

\section{Formalism}

We consider a 5D theory of gravity on which we define a vacuum,
such that the first variation of the action is $^{(5)} \delta{\cal
I}= ^{(5)} \delta{\cal I}_E+^{(5)} \delta{\cal I}_M$, where
\begin{equation}
^{(5)} \delta{\cal I} = \frac{1}{2} \int d^5x\, \sqrt{|g|}\,\delta
g^{ab}\,\left\{\frac{{\cal R}_{ab}}{8\pi G}+ T_{ab} \right\}.
\end{equation}
Here, the first term is the variation of the gravitational
Einstein action $^{(5)} {\cal I}_E$ and the second one is the
variation of the matter action $^{(5)} {\cal I}_M$. Here, $G$ is
the gravitational constant, $g$ is the determinant of the
covariant tensor metric $g_{ab} = \bar{g}_{ab} + \delta
g_{ab}$\footnote{In this letter $a,b$ run from $0$ to $4$ and
Greek letters rum from $0$ to $3$.} and ${\cal R}=\bar{\cal R} +
\delta{\cal R}$ is the Ricci scalar on the metric. Furthermore,
$\bar{g}_{ab}$\footnote{Greek indices run from $0$ to $3$, and
arabic ones from $0$ from $4$.} is the background tensor metric,
which we shall consider as describing a Ricci-flat spacetime. The
energy-momentum tensor of matter, $ T_{ab}$ is defined from the
variation of the matter action $^{(5)} {\cal I}_M$ under a change
of the metric, and will be considered as null to describe the 5D
apparent vacuum. In what follows we shall take in mind a
particular class of background metrics named canonical\cite{mas}
\begin{equation}\label{m1}
d\bar{S}^2= \frac{\psi^2}{\psi^2_0} d\bar{s}^2 -d\psi^2,
\end{equation}
where $d\bar{s}^2 =
\bar{g}_{\alpha\beta}(x^{\alpha},\psi)\,dx^{\alpha} dx^{\beta}$
and $\psi_0$ is introduced to preserve the physical dimensions.
The perturbations, which are in principle very small, are $\delta
g_{ab}=h_{ab}$. Since the fluctuations of the metric are small,
the perturbed Riemann tensor can be linearized
\begin{equation}
\delta{\cal R}_{abcd} \simeq -\frac{1}{2} \left[ h_{bd,ac} +
h_{ac,bd} - h_{bc,ad} - h_{ad,bc} \right],
\end{equation}
so that the Ricci tensor (which is equal to the perturbed Ricci
tensor because the background Ricci tensor will be considered as
zero) $\delta {\cal R}_{ab}= \bar{g}^{cd} \delta{\cal R}_{cabd}$
\begin{equation}\label{ricci}
\delta {\cal R}_{ab} \simeq -\frac{1}{2} \left[ h^c_{\,\,a,cb} +
h^c_{\,\,b,ac} - \bar{g}^{ce} h_{ce,ab} - \Box h_{ab} \right],
\end{equation}
where commas denote the partial derivative and $\Box$ is the
D'Alambertian on the background 5D spacetime.

\subsection{5D Gauge invariant field}

Now we consider the gauge invariant field $\Psi^a_{\,\,b} =
h^a_{\,\,b} - \frac{1}{2} \delta^a_{\,\,b} h$, where $h$ is the
scalar $h= \bar{g}^{ab} h_{ab}$, such that we impose the 5D gauge
\begin{equation}
\Psi^a_{\,\,b;a} =0,
\end{equation}
where the semicolon denotes the covariant derivative. This gauge
implies that $ h^a_{\,\,b,a}= \frac{1}{2} h_{,b} -
\bar{\Gamma}^a_{ca} h^c_{\,\,b} - \bar{\Gamma}^c_{ba}
h^a_{\,\,c}$, $\bar{\Gamma}^a_{bc}$, being the second kind
Christoffel symbols of the 5D background metric. If we consider an
apparent vacuum  $\bar{\cal R}_{ab}=0$, we obtain
\begin{equation}\label{a}
\Box h_{ab} = \bar{g}^{ce} h_{ce,ab} - \left( h^c_{\,\,a,cb} +
h^c_{\,\,b,ac}\right),
\end{equation}
so that the equation (\ref{a}) becomes
\begin{equation}\label{ond}
\Box h_{ab} = \left( \bar{\Gamma}^c_{ec} h^e_{\,\,a}
\right)_{,b}+\left( \bar{\Gamma}^c_{ea} h^e_{\,\,b} \right)_{,c}-
\left( \bar{\Gamma}^e_{ac} h^c_{\,\,e} \right)_{,b} - \left(
\bar{\Gamma}^e_{ba} h^c_{\,\,e} \right)_{,c},
\end{equation}
which is the linearized wave equation for the tensor field
$h_{ab}$ on any 5D Ricci-flat metric, and
\begin{displaymath}
\Box \equiv \bar{g}^{\alpha\beta} \frac{\partial^2}{\partial
x^{\alpha} \partial x^{\beta}} + 2 \bar{g}^{\alpha\psi}
\frac{\partial^2}{\partial x^{\alpha} \partial \psi} +
\bar{g}^{\psi\psi} \frac{\partial^2}{\partial \psi^2}.
\end{displaymath}
The equation (\ref{ond}) describes the dynamics of $h_{ab}(x^a)$
on the 5D Ricci-flat canonical metric (\ref{m1}). Notice that they
have the form $\Box h_{ab} = {\cal S}_{ab}$. Here, ${\cal S}_{ab}$
describes the interaction of $h_{ab}$ with the background metric,
which manifests itself through the connections
$\bar{\Gamma}^e_{ac}$. Of course ${\cal S}_{ab}$ should be null
for a really 5D Riemann-flat metric [i.e., for a 5D Minkowsky
spacetime].

\subsection{Fluctuations on a 5D apparent vacuum}

In order to study the tensor fluctuations $h_{ab}$, on the
canonical metric (\ref{m1}), we shall consider the Einstein
equations $G_{ab} = - 8\pi G\, T_{ab}$, but written in the
following manner
\begin{equation}\label{otro}
R_{ab} = -8\pi G \, \left[ T_{ab} - \frac{1}{3} g_{ab} T\right],
\end{equation}
where we have considered the expressions $g^{ab} g_{ab} =5$ with
$\bar{g}^{ab} \bar{g}_{ab} =5$\footnote{Both expressions imply
that $h^2 +\bar{g}^{ab} h_{ab}+ h^{ab} \bar{g}_{ab} =0$.}, ${\cal
R}_{ab} = \bar{\cal R}_{ab} + \delta {\cal R}_{ab}$, $T_{ab} =
\bar{T}_{ab} + \delta T_{ab}$, and the expansion $g_{ab} =
\bar{g}_{ab} + h_{ab}$ for the 5D Ricci-flat metric. If we
separate the equations on the background and fluctuations in
(\ref{otro}), one obtains
\begin{eqnarray}
&& \bar{\cal R}_{ab} = -8\pi G \,\left[ \bar{T}_{ab} - \frac{1}{3}
\bar{g}_{ab} \bar{T}\right], \label{e1} \\
&& \delta {\cal R}_{ab} = - 8 \pi G \, \left[ \delta T_{ab} -
\frac{1}{3} \left( \bar{g}_{ab} + h_{ab}\right) \, \delta
T\right]. \label{e2}
\end{eqnarray}
The equation (\ref{e1}) describes the background Einstein
equations on the 5D vacuum and (\ref{e2}) the tensor fluctuations
around (\ref{m1}). By multiplying the equation (\ref{e1}) by
$\bar{g}^{ab}$ and taking into account that the metric is
Ricci-flat, we obtain
\begin{equation}\label{e3}
\bar{T}=0,
\end{equation}
which is a manifestation of the absence of matter on 5D. On the
other hand, from the equation (\ref{e2}), it is possible the
obtain the expression that relates the fluctuations of the Ricci
scalar and the energy-momentum one
\begin{equation}\label{e4}
\delta{\cal R} = \frac{16 \pi G}{3} \, \delta T.
\end{equation}
Now we consider the expression (\ref{ricci}), which is {\it valid
only in a linear approximation}. Hence, the equation (\ref{a})
holds
\begin{equation}
\delta T_{ab} = \frac{1}{3} \left( \bar{g}_{ab} + h_{ab} \right)
\, \delta T, \qquad {\rm or} \qquad \delta T_{ab} = \frac{1}{3}
{g}_{ab} \, \delta T.
\end{equation}
Therefore, from the equation (\ref{e2}) one obtains
\begin{equation}
\delta {\cal R}_{ab} =0, \qquad \delta T_{ab} =0,
\end{equation}
so that, as one expects, ${\cal R}$ and $T$ are 5D invariants.

\section{Dynamics of GW on 4D}

Now we consider the dynamics of $h_{ab}$ on a 4D hypersurface
described by $\bar{g}_{\alpha\beta}(x^{\alpha},\psi)$ evaluated on
a particular foliation on the noncompact fifth coordinate
$\psi=\psi_0$:
\begin{equation}
\left.\bar{g}_{\alpha\beta}(x^{\alpha},\psi)\right|_{\psi=\psi_0}.
\end{equation}
We shall consider that the components
$h_{\psi\psi}=h_{a\psi}=h_{\psi a}=0$. The equation (\ref{a}),
evaluated on this hypersurface can be written as
\begin{eqnarray}
^{(4)} \Box h_{ab} & + &  \left. 2 \bar{g}^{\alpha\psi}
\,\frac{\partial^2 h_{ab}}{\partial x^{\alpha} \partial\psi} +
\bar{g}^{\psi\psi}\,\frac{\partial^2 h_{ab}}{\partial \psi^2
}\right|_{\psi=\psi_0} =\frac{\partial}{\partial\psi}
\left(\bar{\Gamma}^{\psi}_{\alpha a} \,h^{\alpha}_{\,\,b}\right) +
\frac{\partial}{\partial x^{\beta}}
\left(\bar{\Gamma}^{\beta}_{\alpha a}
\,h^{\alpha}_{\,\,b}-\bar{\Gamma}^{\alpha}_{b a}
\,h^{\beta}_{\,\,\alpha}\right)  \nonumber \\
&+& \left. \frac{\partial}{\partial x^b} \left(
\bar{\Gamma}^{\alpha}_{\beta\alpha} \,h^{\beta}_{\,\,a} +
\bar{\Gamma}^{\psi}_{\beta \psi} \,h^{\beta}_{\,\,a} -
\bar{\Gamma}^{\alpha}_{a \beta}
\,h^{\beta}_{\,\,\alpha}\right)\right|_{\psi=\psi_0},
\end{eqnarray}
where we must remember that the connections $\bar{\Gamma}^{a}_{b
c}$ are those of the 5D canonical metric (\ref{m1}). Since we are
interested on the $\mu\nu$ components of $h_{ab}$, we obtain
\begin{eqnarray}
^{(4)} \Box h_{\mu\nu} & + &  \left. 2 \bar{g}^{\alpha\psi}
\,\frac{\partial^2 h_{\mu\nu}}{\partial x^{\alpha} \partial\psi} +
\bar{g}^{\psi\psi}\,\frac{\partial^2 h_{\mu\nu}}{\partial \psi^2
}\right|_{\psi=\psi_0} =\frac{\partial}{\partial\psi}
\left(\bar{\Gamma}^{\psi}_{\alpha \mu}
\,h^{\alpha}_{\,\,\nu}\right) + \frac{\partial}{\partial
x^{\beta}} \left(\bar{\Gamma}^{\beta}_{\alpha \mu}
\,h^{\alpha}_{\,\,\nu}-\bar{\Gamma}^{\alpha}_{\nu\mu}
\,h^{\beta}_{\,\,\alpha}\right)  \nonumber \\
&+& \left. \frac{\partial}{\partial x^{\nu}} \left(
\bar{\Gamma}^{\alpha}_{\beta\alpha} \,h^{\beta}_{\,\,\mu} +
\bar{\Gamma}^{\psi}_{\beta \psi} \,h^{\beta}_{\,\,\mu} -
\bar{\Gamma}^{\alpha}_{\mu \beta}
\,h^{\beta}_{\,\,\alpha}\right)\right|_{\psi=\psi_0}.
\end{eqnarray}
The equations of motion for the remaining components are complied
automatically because $h_{\psi\psi}=h_{a\psi}=h_{\psi a}=0$. If we
take into account that $\bar{g}^{\alpha\psi}=0$ and
$\bar{g}^{\psi\psi}=-1$ , we obtain
\begin{eqnarray}
^{(4)} \Box h_{\mu\nu}  & - & \left.\frac{\partial^2
h_{\mu\nu}}{\partial \psi^2 }\right|_{\psi=\psi_0}
=\frac{\partial}{\partial\psi} \left(\bar{\Gamma}^{\psi}_{\alpha
\mu} \,h^{\alpha}_{\,\,\nu}\right) + \frac{\partial}{\partial
x^{\beta}} \left(\bar{\Gamma}^{\beta}_{\alpha \mu}
\,h^{\alpha}_{\,\,\nu}-\bar{\Gamma}^{\alpha}_{\nu\mu}
\,h^{\beta}_{\,\,\alpha}\right)  \nonumber \\
&+& \left. \frac{\partial}{\partial x^{\nu}} \left(
\bar{\Gamma}^{\alpha}_{\beta\alpha} \,h^{\beta}_{\,\,\mu} +
\bar{\Gamma}^{\psi}_{\beta \psi} \,h^{\beta}_{\,\,\mu} -
\bar{\Gamma}^{\alpha}_{\mu \beta}
\,h^{\beta}_{\,\,\alpha}\right)\right|_{\psi=\psi_0}.
\end{eqnarray}
Since we are dealing with 5D canonical metrics like (\ref{m1}), it
is possible to consider the following separation of variables:
$h_{\mu\nu}(x^{\alpha},\psi) \sim
\Sigma(\psi)\,\tilde{h}_{\mu\nu}(x^{\alpha})$. We obtain the
system of equations
\begin{eqnarray}
&& \left.^{(4)} \Box \tilde{h}_{\mu\nu}  -  m^2_g
\tilde{h}_{\mu\nu}\right|_{\psi=\psi_0} ={\cal
S}_{\mu\nu}(x^{\alpha},\psi_0), \label{ee1}
\\
&&\frac{\partial^2 \Sigma(\psi)}{\partial\psi^2} =
m^2_g\,\Sigma(\psi), \label{ee2}
\end{eqnarray}
where
\begin{equation}\label{ee3}
{\cal S}_{\mu\nu}(x^{\alpha}) =
\left.\frac{\partial}{\partial\psi}
\left(\bar{\Gamma}^{\psi}_{\alpha \mu}
\,\tilde{h}^{\alpha}_{\,\,\nu}\right) + \frac{\partial}{\partial
x^{\beta}} \left(\bar{\Gamma}^{\beta}_{\alpha \mu}
\,\tilde{h}^{\alpha}_{\,\,\nu}-\bar{\Gamma}^{\alpha}_{\nu\mu}
\,\tilde{h}^{\beta}_{\,\,\alpha}\right) + \frac{\partial}{\partial
x^{\nu}} \left( \bar{\Gamma}^{\alpha}_{\beta\alpha}
\,\tilde{h}^{\beta}_{\,\,\mu} + \bar{\Gamma}^{\psi}_{\beta \psi}
\,\tilde{h}^{\beta}_{\,\,\mu} - \bar{\Gamma}^{\alpha}_{\mu \beta}
\,\tilde{h}^{\beta}_{\,\,\alpha}\right)\right|_{\psi=\psi_0}.
\end{equation}
The equation (\ref{ee1}) is the Fierz-Pauli\cite{pf} equation of
motion for massive gravitons with mass $m_g$ and a ${\cal
S}_{\mu\nu}$-source. Notice that $m_g$ is induced by the foliation
$\psi=\psi_0$, but the source becomes from the nonzero connections
$\bar{\Gamma}^{\alpha}_{\beta\gamma}$ of the background canonical
metric (\ref{m1}). It is evident from the equation (\ref{ee3})
that the source ${\cal S}_{\mu\nu}$ is originated in the
interaction of the gravitational waves with the background through
the connections $\bar{\Gamma}^{\alpha}_{\beta \gamma}$ of the 5D
background metric (\ref{m1}) evaluated in the static foliation
$\psi=\psi_0$. The general solution of (\ref{ee1}) is
\begin{eqnarray}
\tilde{h}_{\mu\nu}(x) &=& \int d^4x'\, \Delta(x-x')\, {\cal
S}_{\mu\nu}(x')
\nonumber \\
& =&\int d^4x'\,\Delta(x-x')\,\left\{\frac{\partial}{\partial
x'^{\beta}}\left(\bar{\Gamma}^\beta_{\alpha\mu}
\tilde{h}^\alpha_{\,\,\nu}\right) -\frac{\partial}{\partial
x'^{\beta}}\left(\bar{\Gamma}^\alpha_{\nu\mu}
\tilde{h}^\beta_{\,\,\alpha}\right)\right. \nonumber
\\
&+& \left.\frac{\partial}{\partial
x'^{\nu}}\left(\bar{\Gamma}^\alpha_{\beta\alpha}
\tilde{h}^\beta_{\,\,\mu}\right)-\frac{\partial}{\partial
x'^{\nu}}\left(\bar{\Gamma}^\alpha_{\mu\beta}
\tilde{h}^\beta_{\,\,\alpha}\right)\right\}, \label{ee4}
\end{eqnarray}
where $\Delta(x-x')$ is the Green function which obeys: $
\left(^{(4)}\Box- m^2_g\right) \Delta(x-x') = \delta^{(4)}(x-x')$.
The effective action due to the interaction of GW with matter will
be
\begin{equation}
{\cal I}_{Int}=4\pi G\,\int d^4x \,\bar{T}^{\mu\nu}(x)
\tilde{h}_{\mu\nu}(x),
\end{equation}
where $\tilde{h}_{\mu\nu}(x)$ is given by (\ref{ee4}) and the
source ${\cal S}_{\mu\nu}$ by (\ref{ee3}).

\section{Concluding Remarks}

We have studied GW in the framework of the Induced Matter theory
of gravity. In particular, we have restricted our study to
canonical metrics like (\ref{m1}), which are, at least, Ricci-flat
(they could be also Riemann-flat). These kind of metrics are
suitable to describe a 5D physical vacuum on which GW propagates
freely without interactions. We have defined a 5D gauge invariant
field $\Psi^a_{\,\,b}$, to obtain the linearized equations
(\ref{ond}) that describe gravitational waves (for massless
gravitons) on a canonical Ricci-flat metric (\ref{m1}). Note that
they take the form $\Box h_{ab} = {\cal S}_{ab}$, where ${\cal
S}_{ab}$ is the interaction of the GW with the background metric,
which should be null for a really 5D Riemann-flat metric [i.e.,
for a 5D Minkowsky spacetime]. After it, we have obtained the
equations of motion for $h_{\mu\nu}$, on a static foliation of the
metric (\ref{m1}). From the relativistic point of view, observers
that move with frames $U^4 \equiv {d x^4 \over dS}=0$ (described
by a constant foliation on the extra dimension), can see the
massive gravitons moving on a curved hypersurface, such that their
equation of motion is described by the effective 4D Einstein's
equations (\ref{ee1}), with a non-zero fluctuations of the
energy-momentum tensor: ${\cal S}_{\mu\nu}$ given by (\ref{ee3}).
Such a 4D hypersurface is embedded in the 5D apparent vacuum,
which is geometrically described by a 5D Ricci-flat spacetime.
From the mathematical point of view, the Campbell-Magaard
theorem\cite{campbell} serves as a ladder to go between manifolds
whose dimensionality differs by one. This theorem, which is valid
in any number of dimensions, implies that every solution of the 4D
Einstein equations with arbitrary energy-momentum tensor can be
embedded, at least locally, in a solution of the 5D Einstein field
equations in vacuum.

The interesting of the equation (\ref{ee1}) is that describes
gravitons with mass $m_g$ on a 4D curved spacetime, where ${\cal
S}_{\mu\nu}$ represents the interaction of GW with the connections
of the background 5D metric (\ref{m1}) on $\psi=\psi_0$. This is
the main difference with the original Fierz-Pauli formalism, where
${\cal S}_{\mu\nu}$ becomes from the interaction of GW with matter
described by $\bar{T}_{\mu\nu}$\cite{maggiore}.

\section*{Acknowledgements}

\noindent M.B. acknowledges UNMdP and CONICET (Argentina) for
financial support.


\bigskip

\begin{thebibliography}{99}
\bibitem{ei} A. Einstein, {\em Zur allgemeinen Relativit¨atstheorie},
Sitzungsber. Preuss. Akad. Wiss.: 778 (1915).
\bibitem{gr} C. Will. Liv. Rev. Rel. {\bf 4}: 4 (2001); \\
E. Elizalde, S. Nojiri, and S.D. Odintsov. Phys. Rev. {\bf D70}:
043539 (2004); \\
T.P. Sotiriou and V. Faraoni. Rev. Mod. Phys. {\bf 82}: 451
(2010); \\
G. Cognola, E. Elizalde, S. Nojiri, S.D. Odintsov and S. Zerbini.
JCAP {\bf 0502}: 010 (2005); \\
G. Allemandi, A. Borowiec, M. Francaviglia, S. D. Odintsov. Phys.
Rev. {\bf D72}: 063505(2005); \\
S. Nojiri and S. D. Odintsov. Phys. Lett. {\bf B657}: 238(2008);\\
S. Nojiri and S.D. Odintsov. Int. J. Geom. Meth. Mod. Phys.{\bf
4}: 115(2007);\\
S. Capozziello and M. Francaviglia. Gen. Rel. Grav. {\bf 40}: 357
(2008);\\
K. Bamba, S. Nojiri and S. D. Odintsov. JCAP {\bf 0810}:
045(2008).
\bibitem{maggiore} Michele Maggiore, {\em Gravitational Waves}.
Vol. 1: theory and experiments. Oxford University Press. Great
Britain (2008).
\bibitem{bard} J. Bardeen, Phys. Rev. {\bf D22}: 1882 (1980).
\bibitem{guth} A. H. Guth, Phys. Rev. {\bf D23}: 347 (1981);\\
A. A. Starobinsky, in Current Topics in Field Theory, Quantum
Gravity, and Strings, Lecture Notes in Physics Vol. 226 Springer,
New York, 1986.
\bibitem{bell} M. Bellini, Nucl. Phys.{\bf B660}: 389 (2003), Erratum-ibid. {\bf B671}: 483
(2003); \\
D. S. Ledesma, M. Bellini, Phys. Lett. {\bf B581}:1 (2004); \\
J. E. Madriz Aguilar, M. Bellini, Phys. Lett. {\bf B596}: 116
(2004). M. Anabitarte, J. E. Madriz Aguilar, M. Bellini, Eur.
Phys. J. {\bf C45}: 249 (2006);\\
M. Anabitarte, M. Bellini, J. Math. Phys. {\bf 47}: 042502 (2006);
\\
M. Bellini, Phys. Lett. {\bf B632}: 610 (2006).
\bibitem{gwi}
T. L. Smith, H. V. Peiris and A. Cooray, Phys. Rev. {\bf D73}:
123503 (2006);\\
L. P. Grinshchuk. Sov. Phys. JETP {\bf 40}: 409(1975); \\
B. Allen, Phys. Rev. {\bf D37}: 2078 (1988).
\bibitem{im} P. S. Wesson. Space Science Reviews, {\bf 59}: 365
(1992);\\
J. Ponce de Leon. Mod. Phys. Lett. {\bf A16}: 1405 (2001);\\
S. S. Seahra and P. S. Wesson. Class. Quant. Grav. {\bf 19}: 1139
(2002).
\bibitem{wo} J. M. Overduin and P. S. Wesson, Phys. Rept. {\bf 283}: 303
(1997).
\bibitem{mas} B. Mashhoon, H. Liu, P. S. Wesson, Phys. Lett. {\bf
B331}: 305 (1994).
\bibitem{campbell} J. E. Campbell, {\em A course of Differential
Geometry} (Charendon, Oxford, 1926);\\
L. Magaard, {\em Zur einbettung riemannscher Raume in
Einstein-Raume und konformeuclidische Raume}. (PhD Thesis, Kiel,
1963);\\
S. Rippl, C. Romero, R. Tavakol, Class. Quant. Grav. {\bf 12}:
2411 (1995);\\
F. Dahia, C. Romero, J. Math. Phys.{\bf 43}: 5804 (2002);\\
F. Dahia, C. Romero, Class. Quant. Grav. {\bf 22}: 5005 (2005).
\bibitem{pf} M. Fierz, W. Pauli, Proc. Roy. Soc. Lond. {\bf A173}: 211
(1939).
\end{thebibliography}
\end{document}